\documentclass[useAMS,usenatbib]{mn2e}

\usepackage[dvips]{graphicx}

\title[Spectrum Analysis of Bright {\it Kepler} $\gamma$ Doradus Candidate Stars]{Spectrum Analysis of Bright {\it Kepler} $\gamma$ Doradus Candidate Stars\thanks{Based
on observations with the 2-m Alfred-Jensch-Telescope of the
Th\"uringer Landessternwarte Tautenburg}}
\author[A. Tkachenko et al.]
  {A.~Tkachenko,$^1$
  H.~Lehmann,$^2$ B.~Smalley,$^3$ J.~Debosscher,$^1$
  and C. Aerts$^{1,4}$ \\
  $^1$Instituut voor Sterrenkunde, K.U. Leuven, Celestijnenlaan 200D, B-3001 Leuven, Belgium\\
  $^2$Th\"{u}ringer Landessternwarte Tautenburg, 07778 Tautenburg, Germany\\
  $^3$Astrophysics Group, Keele University, Staffordshire ST5 5BG, United Kingdom\\
  $^4$Department of Astrophysics, IMAPP, Radboud University Nijmegen, 6500 GL Nijmegen, The Netherlands}
\date{Received date; accepted date}

\pagerange{\pageref{firstpage}--\pageref{lastpage}} \pubyear{2011}

\def\LaTeX{L\kern-.36em\raise.3ex\hbox{a}\kern-.15em
    T\kern-.1667em\lower.7ex\hbox{E}\kern-.125emX}

\newcommand{\GD}{$\gamma$~Dor}
\newcommand{\DSct}{$\delta$~Sct}
\newcommand{\LBoo}{$\lambda$~Boo}
\newcommand{\vsini}{$v\sin{i}$}

\newcommand{\te}{$T_{\rm eff}$}
\newcommand{\logg}{$\log{g}$}
\newcommand{\kms}{km\,s$^{-1}$}

\begin{document}

\label{firstpage}

\maketitle

\begin{abstract}
 Ground-based spectroscopic follow-up observations of the
pulsating stars observed by the $Kepler$ satellite mission are
needed for their asteroseismic modelling. We aim to derive the
fundamental parameters for a sample of 26 $\gamma$ Doradus
candidate stars observed by the $Kepler$ satellite mission to
accomplish one of the required preconditions for their
asteroseismic modelling and to compare our results with the types
of pulsators expected from the existing light curve analysis. We
use the spectrum synthesis method to derive the fundamental
parameters like \te, \logg, $[M/H]$, and \vsini\ from newly
obtained spectra and compute the spectral energy distribution from
literature photometry to get an independent measure of \te. We
find that most of the derived \te\ values agree with the values
given in the Kepler Input Catalogue. According to their positions
in the HR-diagram three stars are expected \GD\ stars, ten stars
are expected \DSct\ stars, and seven stars are possibly \DSct\
stars at the hot border of the instability strip. Four stars in
our sample are found to be spectroscopic binary candidates and
four stars have very low metallicity where two show about solar C
abundance. Six of the 10 stars located in the \DSct\ instability
region of the HR-diagram show both \DSct\ and \GD-type
oscillations in their light curves implying that \GD-like
oscillations are much more common among the \DSct\ stars than
predicted by theory. Moreover, seven stars showing periods in the
\DSct\ and the \DSct-\GD\ range in their light curves are located
in the HR-diagram left of the blue edge of the theoretical \DSct\
instability strip. The consistency of these findings with recent
investigations based on high-quality $Kepler$ data implies the
need for a revision of the theoretical \GD\ and \DSct\ instability
strips.
\end{abstract}


\section{Introduction}

The {\it Kepler} satellite was launched in March 2009 with the
primary goal to search for transiting exoplanets in the solar
neighbourhood. It delivers single band-pass light curves of
micromagnitude precision and has found hundreds of planet
candidates \citep{Borucki}. The long, uninterrupted, and
high-precision time series photometry taken for a huge number of
stars has also led to the discovery of many new pulsating stars
and is an ideal basis for an in-depth asteroseismic analysis
\citep[see e.g.][]{Gilliland2010}. For this analysis and the
subsequent asteroseismic modelling, precise knowledge of the
fundamental parameters of the stars is essential. These parameters
cannot be determined from the single band-pass photometry
delivered by {\it Kepler} alone, however. Hence, ground-based
spectroscopic follow-up observations have been undertaken to
determine the stellar parameters.

In this paper, we are concerned with $\gamma$~Doradus candidates
found from data assembled with  the {\it Kepler} mission. Such
pulsators are named after their prototype, Gamma Doradus, whose
multiperiodic variable nature was first reported by
\citet{Cousins1992}. \citet{Krisciunas1993} discovered a
multiperiodic photometric variability with an amplitude of about
0.1~mag and periods of 2.275 and 1.277~d in 9~Aurigae and reported
on similar behaviour in $\gamma$~Doradus and HD\,96008. Following
these discoveries, \citet{Balona1994} introduced a new class of
variable stars named after the prototype star $\gamma$ Doradus.
\GD\ stars are assumed to pulsate in high-order, low-degree
non-radial gravity modes driven by the flux blocking mechanism
near the base of their convective zones
\citep{Guzik2000,Dupret2005}. The typical masses of these stars
lie in the range of 1.5--1.8~M$_{\odot}$ \citep{Aerts2010}.
According to \citet{Kaye1999}, \GD-type stars can be characterised
as follows: (1) spectral type A7--F5 and luminosity class IV,
IV-V, or V, (2) low-amplitude photometric variations with periods
between 0.5 and 3 days as well as spectroscopic variability seen
as both line-profile and low-amplitude radial velocity (RV)
variations.

\begin{table} \tabcolsep 1.9mm\caption{\small Journal of observations. N gives the number of obtained spectra, V - the visual magnitude. All spectra have
been taken in 2010.}
\begin{tabular}{rlcrll}
\hline
\multicolumn{1}{c}{KIC\rule{0pt}{9pt}} & \multicolumn{1}{c}{Designation} & $N$ & \multicolumn{1}{c}{$V$} & \multicolumn{1}{c}{SpT} & \multicolumn{1}{c}{Observed}\\
\hline
01\,571\,152\rule{0pt}{9pt} & BD+36\,3535 & 2 & 9.3 & F0 & May-June\\
02\,166\,218 & BD+37\,3490 & 3 & 9.5 & F0 & May-June\\
03\,217\,554 & BD+38\,3415 & 2 & 9.6 & A5 & September\\
03\,453\,494 & BD+38\,3666 & 3 & 9.6 & A5 & September\\
04\,847\,411 & HD\,\,\,\,225314 & 4 & 9.8 & A7 & June-July\\
05\,088\,308 & HD\,\,\,\,180099 & 1 & 8.7 & F5 & May\\
05\,164\,767 & HD\,\,\,\,175537 & 6 & 7.8 & F0 & April-June\\
05\,446\,068 & BD+40\,3704 & 7 & 9.7 & --- & June-July\\
05\,785\,707 & HD\,\,\,\,181902 & 3 & 9.0 & A & August\\
06\,289\,468 & BD+41\,3389 & 5 & 9.4 & A2 & May\\
06\,509\,175 & BD+41\,3248 & 2 & 10.0 & A2 & August\\
06\,587\,551 & BD+41\,3207 & 3 & 9.8 & A0 & June-July\\
06\,756\,386 & HD\,\,\,\,175939 & 2 & 8.7 & A2 & August\\
07\,748\,238 & HD\,\,\,\,181985 & 2 & 9.5 & A & May-July\\
08\,623\,953 & BD+44\,3134 & 3 & 9.3 & A5 & August\\
08\,738\,244 & HD\,\,\,\,176390 & 2 & 8.1 & A3 & August\\
08\,750\,029 & BD+44\,3113 & 3 & 9.7 & A5 & June\\
09\,413\,057 & BD+45\,2954 & 5 & 9.6 & A2 & May-June\\
09\,764\,965 & HD\,\,\,\,181206 & 2 & 8.8 & A5 & August\\
09\,812\,351 & HD\,\,\,\,174019 & 3 & 7.9 & A0 & April-May\\
10\,119\,517 & TYC\,3544-1245-1 & 3 & 9.9 & --- & August\\
10\,451\,090 & HD\,\,\,\,174789 & 3 & 9.2 & A & August\\
10\,616\,594 & TYC\,3561-971-1 & 3 & 9.8 & --- & August\\
10\,977\,859 & HD\,\,\,\,184333 & 2 & 8.8 & A2 & August\\
11\,498\,538 & HD\,\,\,\,178874 & 2 & 7.3 & F5 & September\\
12\,353\,648 & HD\,\,\,\,234859 & 3 & 9.6 & A2 & August\\
\hline
\end{tabular}
\label{Table:observations}
\end{table}

\GD\ stars are located close to the red edge of the classical
instability strip in the HR-diagram. The theoretical \GD\
instability strip overlaps with the red edge of classical
instability strip where the \DSct\ pulsators are located. While
the low-order p-modes of \DSct\ stars are characterized by short
periods ranging from 18 min to 8 h, typical \GD\ high-order
g-modes have periods of the order of a day
\citep[e.g.,][]{Aerts2010}. Multiperiodicity is found for most of
the \GD\ class members from ground-based photometry
\citep{Henry2007,Cuypers2009} and spectroscopy
\citep[e.g.,][]{Mathias2004,DeCat2006}. Pulsators in the
overlapping region of the \DSct and \GD\ instability strip are
expected to show the two pulsation characteristics, i.e.,
high-order g-modes probing the core and low-order p- and g-modes
probing the outer layers. While the frequency patterns of these
two types of oscillations are in principle easy to distinguish in
the co-rotating frame of reference,  the frequencies start to
overlap in an inertial frame of reference, particularly for the
fast rotators. Moreover, the overall beating patterns are complex
and hard to unravel from interrupted ground-based data. Some
hybrid pulsators were already found previously
\citep[e.g.,][]{Rowe2006,King2007}, but the {\it Kepler} data make
it clear that hybrid pulsators turn out to be numerous, both for
AF-type stars \citep[][]{Uytterhoeven2011b,Balona2011a} and for
B-type stars \citep{Balona2011b}.

\citet{Gray1999} were the first to report a connection between the
$\lambda$~Bootis (\LBoo) type stars and the \GD\ variables. \LBoo\
stars are Pop\,I hydrogen burning metal poor (except of C, N, O,
and S) A-type stars \citep[][]{Paunzen,Paunzen2004} showing
significant underabundances of Fe-peak elements (up to --2~dex
compared to the solar composition). They belong to the class of
non-magnetic, chemically peculiar stars. Up to now, only two
further reports \citep{Sadak2006,Rodrig2007} on a possible
connection between the \LBoo\ stars and \GD-type variability
appeared. A recent analysis of a sample of 18 \GD\, stars
performed by \citet{Bruntt2008} revealed no principal difference
between the abundances of the analysed stars and the chemical
composition of non-pulsating A- and F-type stars.

In this paper, we investigate a sample of 26 among the brighter
stars in the {\it Kepler} field which have been proposed to be
candidates for \GD\, variables \citep{Uytterhoeven2011a}. We aim
to evaluate fundamental stellar parameters like effective
temperature \te, surface gravity \logg, projected rotational
velocity \vsini, and microturbulent velocity $\xi$\, as well as
the chemical composition of the target stars from newly obtained
high-resolution spectra. Based on the derived parameters, we
present a classification of the sample stars according to the
expected type of variability. The derived chemical composition in
turn allows to check for a possible connection between \GD-type
variability and \LBoo-type abundance patterns.

\section{Observations}\label{Section:Observations}

\begin{table*}
\tabcolsep 1.2mm\center\caption{\small Fundamental stellar
parameters. The values labeled with ``K'' are taken from the KIC
and given for comparison. Metallicity values labeled with ``(Fe)''
refer to the derived Fe-abundance.}
\begin{tabular}{lclrllrllll}
\hline \multicolumn{1}{c}{KIC\rule{0pt}{9pt}} &
$T_{\rm{eff}}^{\rm{K}}(K)$ & \logg$^{\rm{K}}$ &
\multicolumn{1}{c}{$[M/H]^{\rm{K}}$} & $T_{\rm{eff}}(K)$ &
\multicolumn{1}{c}{\logg} & \multicolumn{1}{c}{$[M/H]$}
& \vsini\,(km\,s$^{-1}$) & $\xi$\,(\kms) & \multicolumn{1}{c}{SpT$^{\rm{K}}$} & \multicolumn{1}{c}{SpT}\\
\hline
01\,571\,152$^{1)}$\rule{0pt}{11pt} & 7048 & 3.164 & +0.05 & 7065$^{+79}_{-79}$ & 4.46$^{+0.23}_{-0.38}$ & --0.18$^{+0.10}_{-0.10}$ & 90.1$^{+8.3}_{-12.7}$ & 2.03$^{+0.46}_{-0.49}$ & F0.5 III  & F1 V\vspace{1.5mm}\\
02\,166\,218 & 7153 & 3.345 & --0.11 & 7062$^{+56}_{-56}$ &
3.88$^{+0.17}_{-0.21}$ & --0.37$^{+0.07}_{-0.07}$ &
99.7$^{+3.5}_{-3.5}$ &
2.82$^{+0.23}_{-0.26}$ & F0 IV-III & F1 IV\vspace{1.5mm}\\
03\,217\,554$^{1)}$ & 7801 & 3.504 & +0.06 & 7667$^{+66}_{-66}$ & 2.78$^{+0.10}_{-0.09}$ & --1.20(Fe) & 225.5$^{+17.0}_{-17.2}$ & 4.82$^{+0.75}_{-0.74}$ & A6.5 IV-III & A7.5 III-II\vspace{1.5mm}\\
03\,453\,494 & 7806 & 3.843 & --0.33 & 7737$^{+57}_{-57}$ & 3.71$^{+0.21}_{-0.22}$ & --0.95(Fe) & 210.8$^{+14.5}_{-14.5}$ & 3.24$^{+0.66}_{-0.60}$ & A7 IV & A7 IV\vspace{1.5mm}\\
04\,847\,411$^{\rm RV)}$ & 6563 & 4.517 & --1.95 &
7466$^{+50}_{-50}$ & 3.83$^{+0.24}_{-0.24}$ &
--0.52$^{+0.09}_{-0.09}$ & 139.9$^{+6.4}_{-6.2}$ &
2.62$^{+0.33}_{-0.33}$ & F4 V  & A8.5 V-IV\vspace{1.5mm}\\
05\,088\,308$^{1)}$ & 6567 & 4.035 & --0.68 & 6708$^{+37}_{-37}$ &
2.67$^{+0.13}_{-0.14}$ & --0.35$^{+0.06}_{-0.07}$ &
40.7$^{+1.2}_{-1.2}$ &
4.01$^{+0.17}_{-0.17}$ & F4 V-IV & F4 III-II\vspace{1.5mm}\\
05\,164\,767$^{\rm RV)}$ & ------ & ------ & ------ & 6933$^{+76}_{-76}$ & 3.59$^{+0.39}_{-0.32}$ & --0.19$^{+0.11}_{-0.12}$ & 163.9$^{+8.7}_{-8.6}$ & 2.62$^{+0.47}_{-0.41}$ & -----------& F1.5 IV\vspace{1.5mm}\\
05\,446\,068$^{1,2)}$ & 5337 & 4.476 & --0.69 & 5763$^{+90}_{-90}$ & 3.37$^{+0.25}_{-0.26}$ & +0.24$^{+0.10}_{-0.10}$ & 7.8$^{+2.1}_{-2.2}$ & 0.0$^{\rm{fixed}}$ & G9.5 V & F9.5 IV\vspace{1.5mm}\\
05\,785\,707 & 8009 & 3.615 & --0.14 & 7965$^{+70}_{-70}$ & 3.37$^{+0.15}_{-0.09}$ & --0.56$^{+0.11}_{-0.11}$ & 171.3$^{+10.4}_{-10.0}$ & 2.84$^{+0.45}_{-0.48}$ & A5.5 IV & A6 IV-III\vspace{1.5mm}\\
06\,289\,468$^{\rm RV)}$ & 8267 & 3.741 & --0.30 &
8107$^{+70}_{-70}$ & 3.30$^{+0.06}_{-0.06}$ &
--0.48$^{+0.11}_{-0.10}$ & 149.7$^{+7.0}_{-7.4}$ &
2.67$^{+0.41}_{-0.39}$ & A4.5 IV & A5 III\vspace{1.5mm}\\
06\,509\,175 & 7299 & 3.522 & --0.21 & 7510$^{+50}_{-50}$ & 3.20$^{+0.25}_{-0.26}$ & --0.45$^{+0.10}_{-0.10}$ & 132.4$^{+7.8}_{-7.2}$ & 2.81$^{+0.43}_{-0.39}$ & A9 IV-III & A8 III\vspace{1.5mm}\\
06\,587\,551 & 8377 & 3.929 & --0.07 & 8826$^{+144}_{-144}$ & 3.76$^{+0.07}_{-0.07}$ & --0.11$^{+0.12}_{-0.12}$ & 139.8$^{+7.9}_{-8.1}$ & 2.80$^{+0.53}_{-0.90}$ & A4 V-IV & A2.5 IV\vspace{1.5mm}\\
06\,756\,386 & 7992 & 3.513 & --0.51 & 7891$^{+62}_{-62}$ & 3.19$^{+0.09}_{-0.08}$ & --0.59$^{+0.10}_{-0.10}$ & 192.8$^{+11.6}_{-11.8}$ & 2.99$^{+0.52}_{-0.49}$ & A6 IV-III & A6 III\vspace{1.5mm}\\
07\,748\,238 & 7228 & 3.470 & --0.22 & 7264$^{+58}_{-58}$ & 3.96$^{+0.20}_{-0.25}$ & --0.37$^{+0.08}_{-0.08}$ & 120.8$^{+4.8}_{-4.9}$ & 3.53$^{+0.33}_{-0.38}$ & A9.5 IV-III & A9.5 V-IV\vspace{1.5mm}\\
08\,623\,953 & 7725 & 3.738 & --0.11 & 7726$^{+50}_{-50}$ & 3.43$^{+0.18}_{-0.17}$ & --0.35$^{+0.08}_{-0.08}$ & 84.8$^{+3.2}_{-3.3}$ & 2.76$^{+0.29}_{-0.29}$ & A7 IV & A7 IV-III\vspace{1.5mm}\\
08\,738\,244 & 8167 & 4.152 & +0.41 & 8154$^{+96}_{-96}$ & 3.24$^{+0.07}_{-0.08}$ & --0.27$^{+0.12}_{-0.12}$ & 133.9$^{+7.1}_{-7.1}$ & 2.70$^{+0.42}_{-0.67}$ & A5 V-IV & A5 III\vspace{1.5mm}\\
08\,750\,029 & ------ & ------ & ------ & 7341$^{+59}_{-59}$ &
3.70$^{+0.27}_{-0.23}$ & --0.56$^{+0.10}_{-0.10}$ &
166.1$^{+8.8}_{-8.4}$ &
2.95$^{+0.46}_{-0.41}$& -----------& A9 IV\vspace{1.5mm}\\
09\,413\,057$^{\rm RV)}$ & 8465 & 3.868 & --0.06 &
8588$^{+97}_{-97}$ & 3.59$^{+0.05}_{-0.05}$ &
--0.56$^{+0.15}_{-0.12}$ & 171.0$^{+10.6}_{-10.4}$ &
2.83$^{+0.57}_{-0.60}$ & A4 V-IV & A3 IV\vspace{1.5mm}\\
09\,764\,965 & 7455 & 4.085 & --0.18 & 7478$^{+41}_{-41}$ & 3.74$^{+0.17}_{-0.18}$ & --0.27$^{+0.06}_{-0.06}$ & 85.1$^{+2.5}_{-2.5}$ & 3.55$^{+0.24}_{-0.24}$ & A8.5 V-IV & A8 IV\vspace{1.5mm}\\
09\,812\,351 & 7794 & 3.470 & --0.29 & 7833$^{+62}_{-62}$ & 3.20$^{+0.15}_{-0.11}$ & --0.90(Fe) & 55.6$^{+3.5}_{-3.2}$ & 2.18$^{+0.40}_{-0.32}$ & A6.5 IV-III & A6 III\vspace{1.5mm}\\
10\,119\,517$^{\rm RV)}$ & 6225 & 4.375 & --0.45 & 6438$^{+69}_{-69}$ & 4.21$^{+0.22}_{-0.13}$ & --0.24$^{+0.07}_{-0.07}$ & 77.9$^{+3.4}_{-3.3}$ & 1.26$^{+0.30}_{-0.27}$ & F8 V & F5 V\vspace{1.5mm}\\
10\,451\,090 & 7577 & 4.134 & --0.05 & 7633$^{+50}_{-50}$ & 3.58$^{+0.15}_{-0.15}$ & +0.04$^{+0.06}_{-0.06}$ & 44.2$^{+1.5}_{-1.5}$ & 3.00$^{+0.16}_{-0.16}$ & A8 V-IV & A7.5 IV-III\vspace{1.5mm}\\
10\,616\,594$^{2)}$ & 5161 & 3.762 & --0.75 & 5327$^{+88}_{-88}$ & 2.81$^{+0.25}_{-0.22}$ & +0.16$^{+0.13}_{-0.20}$ & 7.24$^{+0.9}_{-0.8}$ & 0.63$^{+0.20}_{-0.25}$ & G8.5 V-IV & G3.5 III \vspace{1.5mm}\\
10\,977\,859 & 8052 & 3.935 & +0.15 & 8195$^{+57}_{-57}$ & 3.60$^{+0.07}_{-0.06}$ & --0.11$^{+0.07}_{-0.06}$ & 63.0$^{+3.0}_{-2.9}$ & 3.08$^{+0.28}_{-0.27}$ & A5.5 V-IV & A5 IV\vspace{1.5mm}\\
11\,498\,538$^{2)}$ & 6287 & 4.036 & --0.29 & 6428$^{+57}_{-57}$ & 2.90$^{+0.14}_{-0.15}$ & --0.15$^{+0.07}_{-0.07}$ & 39.7$^{+1.2}_{-1.1}$ & 2.54$^{+0.19}_{-0.16}$ & F7 V-IV & F5.5 III\vspace{1.5mm}\\
12\,353\,648 & 7414 & 3.473 & --0.38 & 7163$^{+51}_{-51}$ & 3.49$^{+0.23}_{-0.29}$ & --1.05(Fe) & 192.0$^{+10.6}_{-10.6}$ & 2.32$^{+0.42}_{-0.43}$ & A8.5 IV-III & F0 IV-III\vspace{1.5mm}\\
\hline \multicolumn{11}{l}{$^{1)}$ suspected SB2 star; $^{2)}$ no
reliable fit obtained; $^{\rm RV)}$ differences in the measured
RVs are observed\rule{0pt}{11pt}}
\end{tabular}
\label{Table:FundamentalParameters}
\end{table*}

\begin{table*}
\tabcolsep 0.9mm\center\caption{\small Derived metallicity and
elemental abundances relative to solar ones in dex. Asterisks
indicate elements with error estimates of $\pm$0.10~dex
($\pm$0.20~dex in all other cases). Metallicity values labeled
with ``(Fe)'' refer to the derived Fe-abundance.} \tiny
\begin{tabular}{cclllllllllllllllll}
\hline
KIC\rule{0pt}{9pt} & $[M/H]$ & \multicolumn{1}{c}{C} & \multicolumn{1}{c}{O} & \multicolumn{1}{c}{Mg} & \multicolumn{1}{c}{Si} & \multicolumn{1}{c}{Ca} & \multicolumn{1}{c}{Fe} & \multicolumn{1}{c}{Na} & \multicolumn{1}{c}{Sc} & \multicolumn{1}{c}{Ti} & \multicolumn{1}{c}{Cr} & \multicolumn{1}{c}{Mn} & \multicolumn{1}{c}{Y} & \multicolumn{1}{c}{Ba} & \multicolumn{1}{c}{V} & \multicolumn{1}{c}{Co} & \multicolumn{1}{c}{Ni} & \multicolumn{1}{c}{Zr}\\
    &            & --3.65 & --3.38 & --4.51 & --4.53 & --5.73 & --4.59 & --5.87 & --8.99 & --7.14 & --6.40 & --6.65 & --9.83 & --9.87 & --8.04 & --7.12 & --5.81 & --9.45\\
\hline
01\,571\,152\rule{0pt}{11pt} & --0.18$^{+0.10}_{-0.10}$ & +0.05 & --- & --0.10$^*$ & --0.10 & --0.25$^*$ & --0.15$^*$ & +0.00 & +0.00 & --0.25$^*$ & --0.15$^*$ & --0.15 & --0.20 & +0.80 & --- & --- & --0.20$^*$ & ---\vspace{2.0mm}\\
02\,166\,218 & --0.37$^{+0.07}_{-0.07}$ & --0.10 & --- & --0.15$^*$ & --0.20 & --0.20$^*$ & --0.45$^*$ & --0.10 & --0.35 & --0.40$^*$ & --0.40$^*$ & --0.40 & --0.35 & +0.60 & --- & --- & --0.45$^*$ & ---\vspace{2.0mm}\\
03\,217\,554 & --1.20(Fe) & --1.10 & --- & --0.35$^*$ & --0.10 & --1.35 & --1.20$^*$ & --- & --1.15 & --1.05 & --1.70 & --- & --- & --1.10 & --- & --- & --0.80 & ---\vspace{2.0mm}\\
03\,453\,494 & --0.95(Fe) & --0.60 & --- & --0.10$^*$ & --0.10 & --1.25 & --0.95$^*$ & --- & --0.75 & --0.95 & --1.00 & --- & --- & --1.05 & --- & --- & --0.70 & ---\vspace{2.0mm}\\
04\,847\,411 & --0.52$^{+0.09}_{-0.09}$ & --0.20 & --- & --0.20$^*$ & --0.10 & --0.70$^*$ & --0.55$^*$ & +0.05 & --0.45 & --0.75$^*$ & --0.55$^*$ & --0.40 & --- & +0.25 & --- & --- & --0.50$^*$ & ---\vspace{2.0mm}\\
05\,088\,308 & --0.35$^{+0.06}_{-0.07}$ & --0.30 & --- & --0.30$^*$ & --0.50 &  --0.25$^*$ & --0.50$^*$ & --0.30 & --0.35 & --0.55$^*$ & --0.45$^*$ & --0.45 & --0.10 & +0.40 & --- & --- & --0.30$^*$ & ---\vspace{2.0mm}\\
05\,164\,767 & --0.19$^{+0.11}_{-0.12}$ & +0.15 & --- & +0.00$^*$ & +0.15 & --0.20$^*$ & --0.35$^*$ & +0.50 & --0.05 & --0.65$^*$ & --0.10$^*$ & --0.05 & --0.60 & +0.60 & --- & --- & --0.25$^*$ & ---\vspace{2.0mm}\\
05\,446\,068 & +0.24$^{+0.10}_{-0.10}$ & --0.30 & --- & +0.10$^*$ & --0.25$^*$ & --0.05$^*$ & +0.15$^*$ & +0.10 & +0.20$^*$ & +0.30$^*$ & +0.25$^*$ & +0.35$^*$ & +0.25$^*$ & +0.50 & +0.55$^*$ & +0.35$^*$ & +0.10$^*$ & +0.10\vspace{2.0mm}\\
05\,785\,707 & --0.56$^{+0.11}_{-0.11}$ & --0.20 & --- & --0.25$^*$ & +0.15 & --0.50 & --0.50$^*$ & --- & --0.50 & --0.85 & --0.70 & --- & --- & --0.40 & --- & --- & --0.45 & ---\vspace{2.0mm}\\
06\,289\,468 & --0.48$^{+0.11}_{-0.10}$ & --0.30 & --- & +0.05$^*$ & --0.05 & --0.75 & --0.50$^*$ & --- & --0.55 & --0.65 & --0.60 & --- & --- & --0.50 & --- & --- & --0.55 & ---\vspace{2.0mm}\\
06\,509\,175 & --0.45$^{+0.10}_{-0.10}$ & --0.15 & --- & --0.05$^*$ & +0.05 & --0.35$^*$ & --0.55$^*$ & +0.10 & --0.45 & --0.65$^*$ & --0.55$^*$ & --0.20 & --- & --0.60 & --- & --- & --0.45$^*$ & ---\vspace{2.0mm}\\
06\,587\,551 & --0.11$^{+0.12}_{-0.12}$ & +0.00 & +0.00 & +0.20$^*$ & +0.30 & --0.20 & --0.10$^*$ & --- & --0.05 & --0.35 & --0.10 & --- & --- & --0.40 & --- & --- & --0.15 & ---\vspace{2.0mm}\\
06\,756\,386 & --0.59$^{+0.12}_{-0.12}$ & --0.20 & --- & +0.00$^*$ & +0.20 & --0.50 & --0.65$^*$ & --- & --0.75 & --0.70 & --0.65 & --- & --- & --1.30 & --- & --- & --0.60 & ---\vspace{2.0mm}\\
07\,748\,238 & --0.37$^{+0.10}_{-0.10}$ & --0.15 & --- & --0.10$^*$ & +0.00 & --0.25$^*$ & --0.45$^*$ & +0.10 & --0.35 & --0.45$^*$ & --0.45$^*$ & --0.35 & --- & +0.50 & --- & --- & --0.40$^*$ & ---\vspace{2.0mm}\\
08\,623\,953 & --0.35$^{+0.10}_{-0.10}$ & --0.10 & --- & --0.10$^*$ & +0.00 & --0.25 & --0.40$^*$ & +0.10 & --0.35 & --0.55 & --0.45 & --0.30 & --- & --0.35 & --- & --- & --0.30 & ---\vspace{2.0mm}\\
08\,738\,244 & --0.27$^{+0.12}_{-0.12}$ & --0.25 & --- & +0.20$^*$ & +0.25 & --0.40 & --0.35$^*$ & --- & --0.50 & --0.60 & --0.20 & --- & --- & --0.55 & --- & --- & --0.35 & ---\vspace{2.0mm}\\
08\,750\,029 & --0.56$^{+0.10}_{-0.10}$ & --0.20 & --- & --0.35$^*$ & +0.00 & --0.60$^*$ & --0.60$^*$ & --0.05 & --0.60 & --0.90$^*$ & --0.50$^*$ & --0.70 & --- & +0.35 & --- & --- & --0.65$^*$ & ---\vspace{2.0mm}\\
09\,413\,057 & --0.56$^{+0.15}_{-0.12}$ & --0.45 & --- & --0.05$^*$ & +0.10 & --0.70 & --0.60$^*$ & --- & --0.85 & --0.75 & --0.65 & --- & --- & --0.30 & --- & --- & --0.60 & ---\vspace{2.0mm}\\
09\,764\,965 & --0.27$^{+0.08}_{-0.08}$ & --0.05 & --- & --0.10$^*$ & +0.00 & --0.15$^*$ & --0.30$^*$ & +0.05 & --0.25 & --0.45$^*$ & --0.30$^*$ & --0.30 & --- & --0.15 & --- & --- & --0.30$^*$ & ---\vspace{2.0mm}\\
09\,812\,351 & --0.90(Fe) & +0.00 & --- & --0.50$^*$ & --0.40 & --0.85 & --0.90$^*$ & --- & --0.85 & --1.00 & --0.85 & --- & --- & --1.20 & --- & --- & --0.65 & ---\vspace{2.0mm}\\
10\,119\,517 & --0.24$^{+0.07}_{-0.07}$ & +0.05 & --- & --0.20$^*$ & --0.15 & --0.05$^*$ & --0.20$^*$ & +0.10 & --0.35 & --0.40$^*$ & --0.20$^*$ & --0.15 & --0.40 & +0.40 & --- & --- & --0.25$^*$ & ---\vspace{2.0mm}\\
10\,451\,090 & +0.04$^{+0.06}_{-0.06}$ & --0.10 & --- & +0.10$^*$ & +0.00 & --0.25 & +0.00$^*$ & +0.25 & --0.15 & --0.15 & +0.15 & +0.00 & --- & +0.75 & --- & --- & +0.20 & ---\vspace{2.0mm}\\
10\,616\,594 & +0.16$^{+0.13}_{-0.20}$ & --0.30 & --- & +0.00$^*$ & --0.35$^*$ & +0.05$^*$ & +0.05$^*$ & +0.10 & +0.10$^*$ & +0.10$^*$ & +0.10$^*$ & +0.20$^*$ & +0.20$^*$ & +0.45 & +0.30$^*$ & +0.05$^*$ & +0.00$^*$ & --0.20\vspace{2.0mm}\\
10\,977\,859 & --0.11$^{+0.07}_{-0.06}$ & --0.05 & --- & +0.20$^*$ & +0.10 & --0.05 & --0.10$^*$ & --- & --0.15 & --0.35 & --0.10 & --- & --- & --0.25 & --- & --- & --0.05 & ---\vspace{2.0mm}\\
11\,498\,538 & --0.15$^{+0.07}_{-0.07}$ & --0.15 & --- & +0.20$^*$ & --0.15 & --0.15$^*$ & --0.20$^*$ & --0.15 & --0.20 & --0.30$^*$ & --0.10$^*$ & --0.15 & --0.40 & +0.65 & --- & --- & --0.20$^*$ & ---\vspace{2.0mm}\\
12\,353\,648 & --1.05(Fe) & --0.25 & --- & --0.55$^*$ & --0.25 & --0.90$^*$ & --1.05$^*$ & --- & --1.35 & --2.05$^*$ & --1.25$^*$ & --- & --- & --0.75 & --- & --- & --1.05$^*$ & ---\vspace{1.5mm}\\
\hline
\end{tabular}
\label{Table:IndividualAbundances}
\end{table*}

We base our analysis on high-resolution, high signal-to-noise
ratio (S/N) spectra taken with the Coud\'{e}-Echelle spectrograph
attached to the 2-m telescope of the Th\"{u}ringer
Landessternwarte Tautenburg, Germany. The spectra have a
resolution of 32000 and cover the wavelength range from 4720 to
7400~\AA. Table~\ref{Table:observations} represents the journal of
observations and gives the Kepler Input Catalog (KIC) number, an
alternative designation, the number of obtained spectra, the
visual magnitude, the spectral type as is indicated in the SIMBAD
database, and the period of observations in 2010. The number of
acquired spectra is different for different stars since we aimed
to reach a S/N of about 100 for the mean, averaged spectrum of
each object.

The data have been reduced using standard ESO-MIDAS packages. The
data reduction included bias and stray-light subtraction, cosmic
rays filtering, flat fielding by a halogen lamp, wavelength
calibration by a ThAr lamp, and normalisation to the local
continuum. All spectra were additionally corrected in wavelength
for individual instrumental shifts by using a large number of
telluric O$_2$ lines. The cross-correlation technique was used to
estimate the RVs from the individual spectra so that the single
spectra finally could be shifted and added to build the mean, high
S/N ratio averaged spectrum of each star. The RVs computed at this
step have also been used to check for possible variations due to
binarity, high-amplitude stellar oscillations and rotational
modulation.

\section{Spectral analysis}

\subsection{Method}\label{Section:Methods}

The ``classical'' method of spectrum analysis by means of
equivalent width measurements and subsequent fitting of the
ionization equilibria of different elements requires the star to
rotate slowly, so that a sufficient number of clean, un-blended
spectral lines can be identified and measured in the stellar
spectrum. In the case of rapidly rotating stars, this method fails
due to the high percentage of blended lines. The method of
spectrum synthesis, on the other hand, is based on the comparison
between observed and theoretical spectra in a certain wavelength
range. Its advantage is that the effect of line blending can be
taken into account when computing the synthetic spectra and thus
no restrictions with respect to the rotational velocity occur.
Because of the large number of fast rotators in our sample, we use
the second method comparing the observed spectra with a huge
number of synthetic spectra computed on a grid in the stellar
parameters. Against the much faster approach of solving the
so-called inverse problem, i.e. to determine the physical
parameter values directly from the observations using some
non-linear optimization method, the grid search has the advantage
that it will always determine the globally best solution if the
grid is dense enough. Its principal disadvantage of much longer
computing time plays no crucial role in our analysis since the
synthetic spectra are calculated using a large library of
pre-computed model atmospheres
(Table\,\ref{Table:AtmosphereModels}) and the analysis runs very
fast on up to 300 processor cores of a cluster PC.

Our code GSSP (Grid Search in Stellar Parameters) finds the
optimum values of effective temperature \te, surface gravity
\logg, microturbulent velocity $\xi$, metallicity $[M/H]$, and
projected rotational velocity \vsini\ from the minimum in $\chi^2$
obtained from a comparison of the observed spectrum with the
synthetic ones computed from all possible combinations of the
before mentioned parameters. The errors of measurement (1$\sigma$
confidence level) are calculated from the $\chi^2$ statistics. A
detailed description of the method is given in \citet{Lehmann2011}
(Paper~I).

For the calculation of synthetic spectra, we use the LTE-based
code SynthV \citep{Tsymbal1996} which allows the computation of
spectra based on individual elemental abundances. The code uses
pre-calculated atmosphere models which have been computed with the
most recent, parallelised version of the LLmodels program
\citep{Shulyak2004}. Both programs make use of the VALD database
\citep{Kupka2000} for a pre-selection of atomic spectral lines.
The main limitation of the LLmodels code is that the models are
well suitable for early and intermediate spectral type stars but
not for very hot and cool stars where non-LTE effects or
absorption in molecular bands may become relevant, respectively.

\begin{table} \tabcolsep 2.3mm\caption{\small $E(B-V)$ determined from the Na\,D lines,
\te\ obtained from SED-fitting, and the reddening-corrected \te.}
\begin{tabular}{rccc}
\hline
\multicolumn{1}{c}{KIC\rule{0pt}{9pt}} & $E(B-V)$ & \te & \te\ (dered)\\
\hline
01\,571\,152\rule{0pt}{9pt} & 0.03 & 6820$\pm$140 & 6980$\pm$190\\
02\,166\,218 & 0.01 & 7050$\pm$140 & 7110$\pm$200\\
03\,217\,554 & 0.04 & 7650$\pm$160 & 7910$\pm$250\\
03\,453\,494 & 0.03 & 7650$\pm$150 & 7840$\pm$240\\
04\,847\,411 & 0.05 & 7290$\pm$150 & 7600$\pm$220\\
05\,088\,308 & 0.02 & 6730$\pm$150 & 6840$\pm$190\\
05\,164\,767 & 0.06 & 6770$\pm$130 & 7100$\pm$190\\
05\,446\,068 & 0.04 & 4950$\pm$110 & 5060$\pm$130\\
05\,785\,707 & 0.03 & 7840$\pm$160 & 8060$\pm$260\\
06\,289\,468 & 0.03 & 8130$\pm$170 & 8300$\pm$280\\
06\,509\,175 & 0.08 & 7080$\pm$140 & 7560$\pm$220\\
06\,587\,551 & 0.05 & 8280$\pm$170 & 8760$\pm$350\\
06\,756\,386 & 0.01 & 7860$\pm$150 & 7930$\pm$240\\
07\,748\,238 & 0.05 & 7150$\pm$150 & 7450$\pm$220\\
08\,623\,953 & 0.04 & 7720$\pm$150 & 7990$\pm$250\\
08\,738\,244 & 0.01 & 8110$\pm$160 & 8190$\pm$240\\
08\,750\,029 & 0.05 & 7250$\pm$140 & 7560$\pm$220\\
09\,413\,057 & 0.05 & 8270$\pm$170 & 8720$\pm$340\\
09\,764\,965 & 0.01 & 7450$\pm$170 & 7510$\pm$230\\
09\,812\,351 & 0.01 & 7750$\pm$150 & 7830$\pm$230\\
10\,119\,517 & 0.00 & 6380$\pm$160 & 6380$\pm$160\\
10\,451\,090 & 0.01 & 7700$\pm$160 & 7750$\pm$240\\
10\,616\,594 & 0.00 & 5240$\pm$140 & 5240$\pm$140\\
10\,977\,859 & 0.01 & 8130$\pm$170 & 8200$\pm$260\\
11\,498\,538 & 0.03 & 6490$\pm$160 & 6530$\pm$190\\
12\,353\,648 & 0.02 & 7350$\pm$180 & 7470$\pm$230\\
\hline
\end{tabular}
\label{Table:SED}
\end{table}

\begin{figure}
\centering
\includegraphics[scale=0.85]{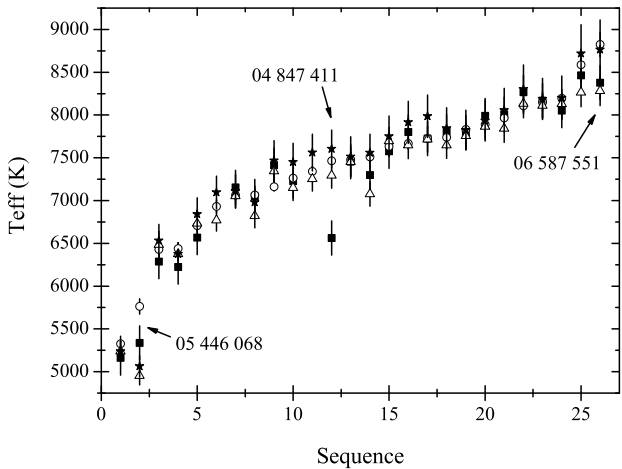}
\caption{{\small Comparison of \te\ derived spectroscopically
(open circles) with the KIC (filled boxes) and photometric values
(open triangles and stars). See text for detailed description.}}
\label{Temperature_comp}
\end{figure}

\begin{table} \tabcolsep 0.0mm\caption{\small
Stellar atmosphere models computed with the LLmodels code for
$\xi$\,=\,2\,\kms.}
\begin{tabular}{lclclc}
\hline\hline
\multicolumn{6}{c}{Parameter, step width\rule{0pt}{9pt}}\\
\hline \multicolumn{1}{c}{$[M/H]$\rule{0pt}{9pt}} & $\Delta[M/H]$
&\multicolumn{1}{c}{\te(K)}
   & $\Delta$\te(K) & \multicolumn{1}{c}{\logg} & $\Delta$\logg\\
\hline --0.8\,--\,+0.8 & 0.1 & \begin{tabular}{l}
                       ~~4\,500\,--\,10\,000\rule{0pt}{9pt}\\ 10\,000\,--\,22\,000\\
                       \end{tabular}
                     & \begin{tabular}{c}
                       100\rule{0pt}{9pt}\\ 250\\
                       \end{tabular}
                     & \begin{tabular}{c}
                       2.5\,--\,5.0\rule{0pt}{9pt}\\ 3.0\,--\,5.0\\
                       \end{tabular}
                     & 0.1\\
\hline
\multicolumn{5}{l}{Total number of models:\rule{0pt}{9pt}} & \textbf{41\,888}\\
\hline
\end{tabular}
\label{Table:AtmosphereModels}
\end{table}

The method has been tested on spectra of Vega and successfully
applied to {\it Kepler} $\beta$\,Cep and SPB candidate stars
(Paper~I). In Paper~I, the chemical composition of the stars has
been determined by means of an iterative procedure involving (1)
the estimation of \te, \logg, \vsini, $\xi$, and $[M/H]$, (2) the
determination of the individual abundances, element-by-element, by
fixing all parameters derived in the previous step and taking the
abundance table corresponding to the derived metallicity as a
first guess, and (3) the re-estimation of \te, \logg, \vsini, and
$\xi$ based on the chemical composition evaluated in the second
step. In the latest version of the GSSP code, we still iterate the
individual abundances element-by-element after the first step, but
together with \te, \logg, \vsini, and $\xi$. This allows us to
avoid the additional third step. In order to allow the possibility
of an incorrect normalisation to be taken into account, and to
minimize its influence on the results, we also introduced an
additional free parameter that allows the adjustment of the
observed continuum relative to the synthetic one during the
fitting procedure.

\subsection{Results}\label{Section:Results}

Table~\ref{Table:FundamentalParameters} summarises the results of
spectrum analysis for all 26 stars of our sample. The first four
columns of the table represent correspondingly the KIC-number of a
star, the effective temperature $T_{\rm{eff}}^{\rm{K}}$, surface
gravity $\log{g}^{\rm{K}}$, and metallicity $[M/H]^{\rm{K}}$ as is
indicated in the KIC. The five following columns list the stellar
parameters derived from our spectra, while the last two columns
represent the spectral types as estimated from \te\ and \logg\
given in the KIC and determined in this work, respectively. In
both cases, the spectral types and the luminosity classes have
been derived using an interpolation in the tables by
\citet{Schmidt-Kaler1982}. We achieve a mean accuracy of about
$1\%$ for \te, about $\pm$ 0.16~dex for \logg, and about $5\%$ for
\vsini.

Table~\ref{Table:IndividualAbundances} lists the elemental
abundances derived for each target star. The metallicity given in
the second column of the table refers to the initially derived
chemical composition and was used as initial guess for the
determination of the individual abundances. All abundances are
given relative to solar values, i.e. negative/positive values
refer to an under-/overabundance of the corresponding element
compared to the solar composition. We assume the chemical
composition of the Sun given by \citet{Grevesse2007} and these
values are listed in the header of
Table~\ref{Table:IndividualAbundances} below the element
designations. For some of the stars we have reached the
metallicity limit in our grid of atmosphere models
(KIC\,03\,217\,554, 03\,453\,494, 09\,812\,351, and 12\,353\,648).
In these cases, we give the derived Fe-abundance instead of the
metallicity. In all other cases, the derived Fe abundance matches
the derived metallicity within the measurement error. The
abundance errors are estimated to be about $\pm$0.1~dex for the
elements showing a sufficient number of strong spectral lines in
the considered region and about $\pm$0.2~dex for the elements
represented in the spectrum by only few and rather faint spectral
lines.

\subsection{Special characteristics of some target stars}

\begin{figure}
\includegraphics[scale=0.8,clip=]{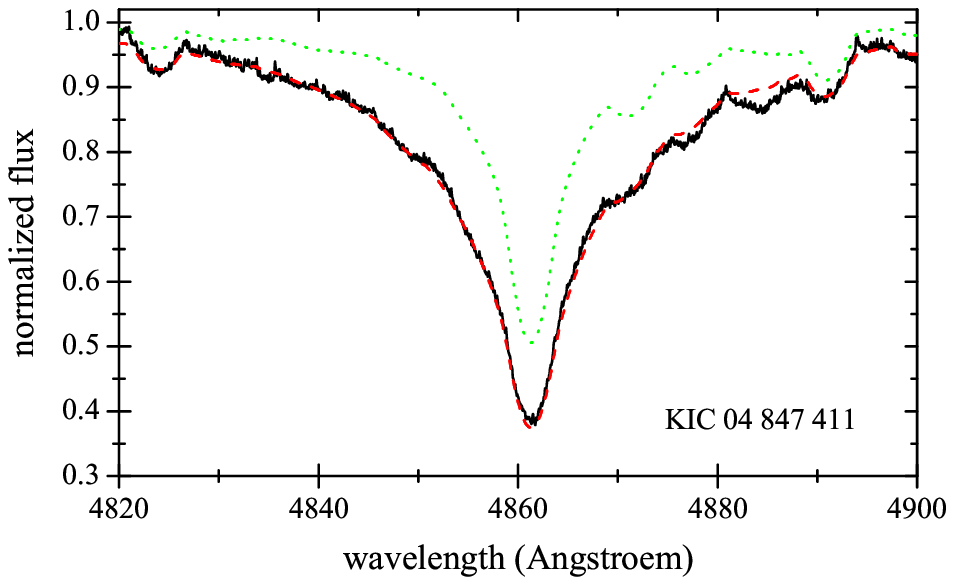}
\includegraphics[scale=0.8,clip=]{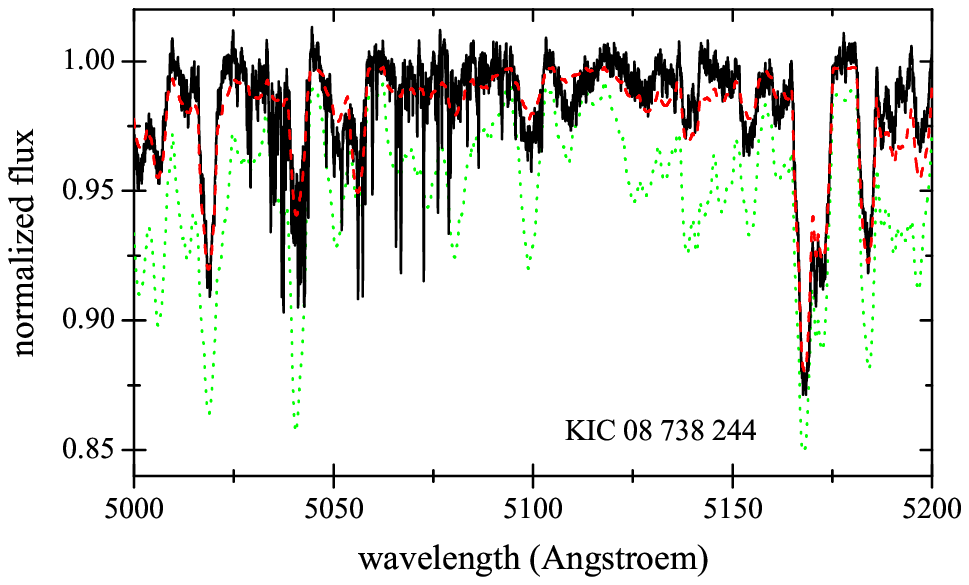}
\includegraphics[scale=0.8,clip=]{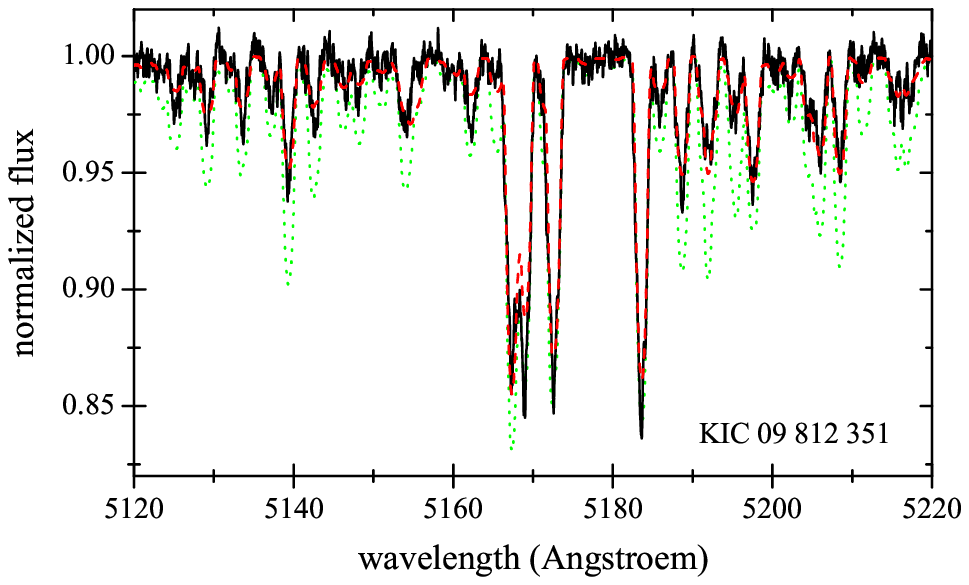}
\includegraphics[scale=0.8,clip=]{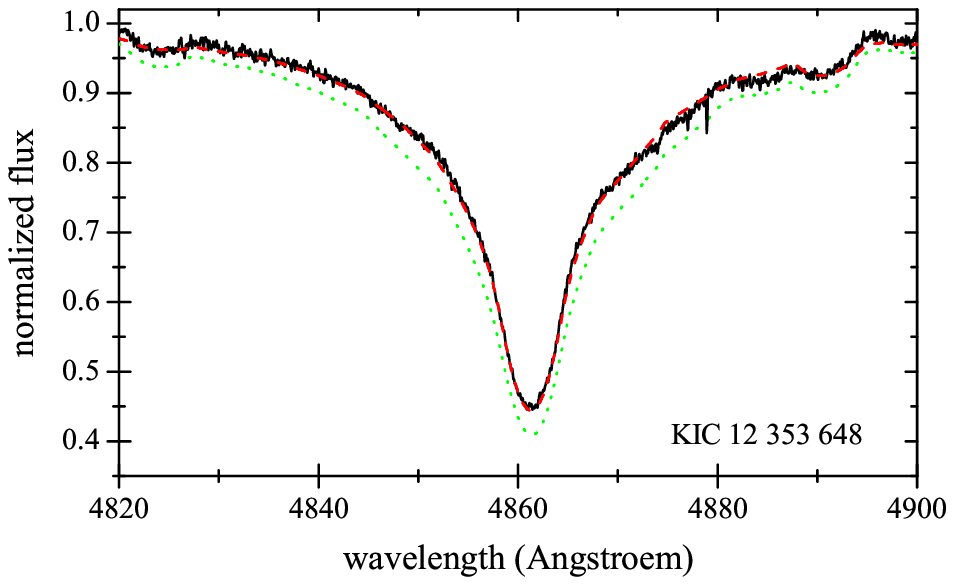}
\caption{{\small Fit of observed spectra (solid, black line) by
synthetic spectra calculated from our optimised parameters
(dashed, red line) and from the values given in the KIC (dotted,
green line), showing either the H$_\beta$ or a metal lines region.
A colour plot is provided in the online version.}}
\label{CompProfiles}
\end{figure}

The derived \te\ and \logg\ are discussed in
Sect.\,\ref{Section:KIC}. Here, we focus on metallicity and
abundance anomalies based on
Table~\ref{Table:IndividualAbundances}, and on possible binarity
of the target stars.

{\it Stars of lower metallicity.} Only three stars in our sample
of 26 show metallicities slightly higher than the Sun; all other
stars have lower metallicity. Fifteen stars have a metallicity of
more than 0.3 dex lower than the Sun. The four stars of lowest
metallicity (KIC\,03217554, 03453494, 09812351, and 12353648) show
underabundances of the Fe-peak elements and Ca of about 1~dex but
much less for Mg and Si. Two of them (KIC\,09\,812\,351 and
12\,353\,648) have C abundances comparable to the solar value,
resembling the characteristics of \LBoo\ stars.

{\it Abundance anomalies.} For eleven of the analysed stars the Ba
abundance is found to deviate by more than 0.4~dex from the
derived metallicity. In only one of them Ba is underabundant, all
other are Ba enhanced. Since the Ba abundance was determined from
only one resonance line at $\lambda$ 4934\,\AA\ which is known to
be strong and sensitive to non-LTE effects, this result must be
interpreted with caution.

\begin{table} \tabcolsep 2.0mm\caption{\small The stars for which remarkable differences in the individual RVs are observed.}
\begin{tabular}{rcrcc}
\hline
\multicolumn{1}{c}{KIC\rule{0pt}{9pt}} & BJD-245000 & \multicolumn{1}{c}{RV} & dRV & max. diff\\
 & & (\kms) & (\kms) & (\kms)\\
\hline
04\,847\,411\rule{0pt}{9pt} & 362.552194 & 2.189 & 0.058 &\\
& 365.476249 & 2.136 & 0.137 &\\
& 381.512639 & --0.968 & 0.204 &\\
& 388.449139 & --3.357 & 0.267 & 5.5\\
05\,164\,767 & 316.464427 & 1.493 & 0.129 &\\
& 316.486187 & 0.798 & 0.152 &\\
& 316.509927 & --1.886 & 0.083 &\\
& 318.457963 & --0.599 & 0.109 &\\
& 318.479550 & 0.194 & 0.123 & 3.4\\
06\,289\,468 & 338.560657 & --0.248 & 0.285 &\\
& 345.482244 & --2.466 & 0.512 &\\
& 345.510347 & --0.107 & 0.368 &\\
& 345.532246 & 0.507 & 0.420 &\\
& 345.554214 & 2.314 & 0.394 & 4.8\\
09\,413\,057 & 340.573022 & --5.208 & 0.104 &\\
& 351.507244 & --6.058 & 0.082 &\\
& 363.430141 & 0.895 & 0.540 &\\
& 365.404034 & 7.092 & 0.736 &\\
& 376.484499 & 3.279 & 0.517 & 13\\
10\,119\,517 & 428.472508 & --110.03 & 0.066 &\\
& 429.421495 & 80.42 & 0.058 &\\
& 430.379879 & 29.61 & 0.059 & 190\\
\hline
\end{tabular}
\label{Table:RVs_indiv}
\end{table}

Most of the stars show overabundances of Mg and Si compared to the
derived metallicity, and for eight stars Na is found to be
significantly enhanced. Additionally, Ti is found to be depleted
in the atmospheres of KIC\,05\,164\,767, 08\,750\,029 and
12\,353\,648. For KIC\,03\,217\,554, we did not find consistent
results for the iron peak elements, there is a large scatter among
the abundances of Fe, Cr, and Ni.

{\it RV variations.} Five stars in our sample show differences in
the measured RVs that are much larger than the errors of
measurement. They are listed in Table~\ref{Table:RVs_indiv} (note
that the RVs are on a relative scale so that the mean RV is zero
for each star). From an inspection of the RVs, we suspect for two
of these stars (KIC\,04\,847\,411 and 09\,413\,057) a periodic
variation - although the time basis is too short to distinguish
between possible periods. KIC\,09\,413\,057 is found by
\citet{Uytterhoeven2011b} to be a hybrid pulsator. One of the
possible periods that fits the observed RV variations of this star
is about 1.6~d and could be caused by pulsations. For one of the
five stars (KIC\,10\,119\,517) we observed an extremely large
difference in RV of 190 \kms\ within one day.

\section{Spectral energy distributions}

The effective temperature can also be determined from the spectral
energy distribution (SED). For our target stars we constructed
SEDs using literature photometry from 2MASS \citep{Skrutskie2006},
Tycho \citep{Hoeg1997}, TASS \citep{Droege2006}, USNO-B1
\citep{Monet2003} and UCAC3 \citep{Zacharias2010}. \te\ was
determined by fitting solar-composition \cite{Kurucz1993} model
fluxes to the photometry. The model fluxes were convolved with the
photometric filter response functions. A weighted
Levenberg-Marquardt non-linear least-squares fitting procedure was
used to find the solution that minimised the difference between
the observed and model fluxes. Since $\log g$ is poorly
constrained due to lack of UV flux measurements, we fixed log\,$g$
= 4.0 for all the fits. This introduces additional errors of about
50~K into the determined effective temperature for the stars
showing significant deviations from the assumed value of \logg.

Since spectral energy distributions can be significantly affected
by interstellar reddening, we measured the equivalent widths of
the interstellar Na\,D lines if present in our spectra and
determined $E(B-V)$ using the relation given by \cite{Munari1997}.
Table~\ref{Table:SED} lists the results. Columns 3 and 4
correspondingly give the temperature values derived without and
with the effects of interstellar reddening taken into account.

\section{Comparison with the Kepler Input Catalogue}\label{Section:KIC}

Figure~\ref{Temperature_comp} compares the spectroscopically
derived \te\ with the photometric and KIC values (typical errors
of the KIC data are $\pm$200~K for \te\ and $\pm$0.5~dex for both
\logg\ and metallicity). The stars are sorted by the spectroscopic
\te\ value starting with the coolest object. For most of the
targets we find a rather good agreement between the
spectroscopically determined temperature (open circles) and the
value listed in the KIC (filled boxes). Whereas in most cases the
\te\ from the uncorrected SED fit (open triangles) is slightly
lower than the spectroscopic one, the temperatures corrected for
the interstellar reddening (asterisks) show rather a good
agreement. In the following, we discuss stars that show larger
deviations from this general tendency in \te\ based on
Fig.\,\ref{Temperature_comp} or large deviations from the KIC
values in the other parameters based on
Table\,\ref{Table:FundamentalParameters}.

{\it KIC\,05\,446\,068:} The spectroscopically derived temperature
exceeds the KIC value by 400\,K and the de-reddened photometric
one by 700\,K. Our fit with the best synthetic spectrum is rather
poor and the spectrum of a second star can clearly be seen in the
residuals. We assume the star to be a SB2 star so that none of the
derived temperatures may be valid.

\begin{figure}
\centering
\includegraphics[scale=0.85]{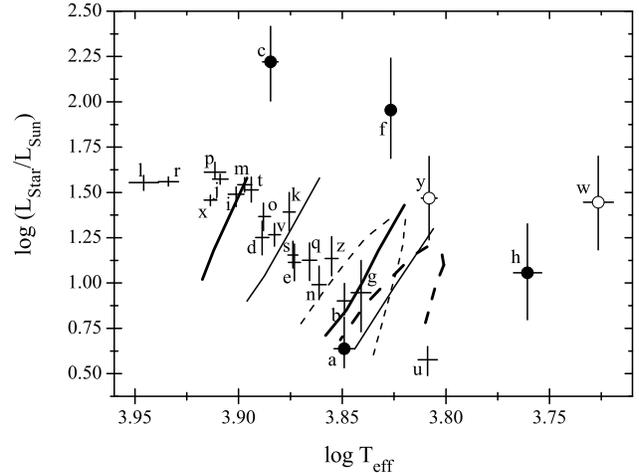}
\caption{Location of the stars (see
Table~\ref{Table:Classification} for labels) and the \GD\ (dashed
lines) and \DSct\ (solid lines) theoretical instability strips in
the HR-diagram. Filled circles indicate suspected binaries, open
circles the stars for which no reliable fit has been obtained.}
\label{HR_diagram}
\end{figure}

\begin{table}
\caption{Classification according to the type of variability
derived from spectrum and light curve analysis.}
\begin{tabular}{|l|c|c|c|}
\hline \multicolumn{1}{|c|}{KIC}               &
spectroscopic\rule{0pt}{9pt}             & light curve &label\\
\hline
01\,571\,152$^{1)}$\rule{0pt}{9pt}    &                                   &                               &  a  \\
02\,166\,218                       & \GD s or hybrids & \GD s &  b \\
05\,164\,767\rule{0pt}{9pt}       & & &  g  \\
\hline
03\,453\,494\rule{0pt}{9pt}       &                                    &                                  &  d  \\
06\,509\,175                   &                                &                        &  k  \\
07\,748\,238                   &                                &  \raisebox{1.5ex}[-1.5ex]{hybrids} &  n  \\
09\,764\,965                   &                                &
&  s  \\\cline{3-4} 04\,847\,411$^{3)}$\rule{0pt}{3pt}       & & &  e  \\
08\,623\,953                   &   \raisebox{1.5ex}[-1.5ex]{\DSct s}                             &                                  &  o  \\
08\,750\,029                   &                                &                        &  q  \\
09\,812\,351                   &                                    &  \raisebox{1.5ex}[-1.5ex]{\DSct s}                                &  t  \\
10\,451\,090                   &                                &                                  &  v  \\
12\,353\,648                   & & &  z  \\ \hline
06\,289\,468\rule{0pt}{9pt}                   &                                &                                  &  j  \\
06\,587\,551       &                                &                                  &  l  \\
06\,756\,386                       &                        & hybrids &  m  \\
08\,738\,244                   & possibly \DSct s & &  p  \\
09\,413\,057                   &                                &
&  r  \\\cline{3-4} 05\,785\,707\rule{0pt}{9pt}                   &            &                                  &  i  \\
10\,977\,859                   & & \raisebox{1.5ex}[-1.5ex]{\DSct s} &  x  \\
\hline
05\,446\,068$^{1,2)}$\rule{0pt}{9pt}  &                                    &hybrid                        &  h  \\
10\,616\,594$^{2,3)}$            & too cool                           & \DSct                        &  w  \\
10\,119\,517                   & &not pulsating &  u  \\
\hline
03\,217\,554$^{1)}$\rule{0pt}{9pt}    &                                    &\DSct                            &  c  \\
05\,088\,308$^{1)}$            & too evolved                                & \GD                        &  f  \\
11\,498\,538$^{2)}$            &                                    &no classification                 &  y  \\
\hline
\multicolumn{4}{l}{$^{1)}$ suspected SB2 star; $^{2)}$ no reliable fit obtained;\rule{0pt}{11pt}}\\
\multicolumn{4}{l}{$^{3)}$ not analyzed by
\citet{Uytterhoeven2011b}}\\
\end{tabular}
\label{Table:Classification}
\end{table}

{\it KIC\,06\,587\,551:} According to the spectroscopic findings,
this is the hottest star of our sample, in agreement with the
photometrically evaluated \te\ corrected for the interstellar
reddening. Both the KIC and the uncorrected photometric values are
by about 500~K lower. This is an interesting fact since we already
showed in Paper~I that, for the hotter stars (\te$>$8000\,K), the
KIC values of \te\ are systematically lower than the spectroscopic
ones, suspecting that the reason may be that the interstellar
reddening was not properly taken into account when deriving the
KIC temperatures.

{\it KIC\,04\,847\,411:} This example shows that the parameters
listed in the KIC can be unreliable. Figure~\ref{CompProfiles}
(top panel) compares the observed H$_\beta$ line profile (solid,
black line) with the synthetic ones computed from the parameters
derived by us (dashed, red line) and from those listed in the KIC
(dotted, green line). Obviously, the green spectrum does not match
the observations at all. Since the metallicity value of $-1.95$
listed in the KIC is much lower than the limiting value in our
grid of atmosphere models, we expect the deviation from the
observations to be even larger, in particular for all metal lines.
Our spectra show RV variations with an amplitude of about
5.5\,\kms\ and a period of $\sim$10~d. We need more spectra to
reveal the nature of this variability.

{\it KIC\,08\,738\,244:} This star shows a large discrepancy
between the derived values of \logg\ and $[M/H]$ and those listed
in the KIC. Figure~\ref{CompProfiles} (second panel) compares the
observed spectrum with synthetic spectra in one metal line region
and shows that the model based on the KIC values gives no reliable
fit. The same is the case for the Balmer line profiles.

{\it KIC\,09\,812\,351 and 12\,353\,648:} these are metal poor
stars with metallicities below the limit of our grid of atmosphere
models. Both show overabundances of C, Mg, and Si compared to the
derived Fe abundance, while the spectrum of KIC\,12\,353\,648
additionally exhibits a strong depletion of Ti. The derived
metallicities, represented in this case by the Fe abundance, are
much lower than the values listed in the KIC and KIC\,12\,353\,648
additionally shows a discrepancy of about 300~K in \te. The effect
of the different parameter sets on the synthetic spectra is
illustrated in two lower panels of Figure~\ref{CompProfiles}.

Figure~\ref{HR_diagram} shows the positions of all stars of our
sample in the log\,(L/L$_{\odot}$)--log\,\te\ diagram, together
with the \DSct\ and \GD\ instability strips. The latter were
reconstructed from \citet[ Figures 2 and 9]{Dupret2005}. The edges
of the \DSct\ instability region have been computed with a
mixing-length parameter of $\alpha$=1.8 for the fundamental mode
(solid thin lines) and for a radial order of $n$=4 (solid thick
lines). The edges of the \GD\ instability regions computed with
$\alpha$=2.0 and 1.5 are represented by dashed thin and thick
lines, respectively. To place the stars into the diagram, we
estimated their luminosities from the spectroscopically derived
\te\ and \logg\ by means of an interpolation in the tables by
\citet{Schmidt-Kaler1982}. The luminosity error bars represent a
combination of the errors in \te\ and \logg, and so in some cases
they appear to be significantly larger than the uncertainties in
\te. Beside that, the luminosity errors can still be
underestimated due to the uncertainties in the empirical
relations. Realizing that, we base our classification mainly on
the position of the stars in the HR-diagram according to the
derived temperatures.


\section{The stars in the HR-diagram}

Table~\ref{Table:Classification} classifies the stars according to
their type of variability, listing the classifications expected
from their location in the HR-diagram (Fig.\,\ref{HR_diagram}) and
derived by \citet{Uytterhoeven2011b} from the frequency analysis
of the {\it Kepler} light curves. There are six ``outliers'' in
the log\,(L/L$_{\odot}$)-log\,\te\ diagram
(Fig.~\ref{HR_diagram}). Three of them (labels c, f, and h) are
suspected binaries and two (labels w and y) are the stars for
which no reliable fit of the observed spectrum could be obtained.
For these objects we cannot give a certain classification. For the
remaining, sixth object, KIC\,10\,119\,517 (label u), no
pulsations could be found from the light curve analysis.

For most stars of our sample, the classification based on the
light curve analysis appears to be fully consistent with the
position of the objects in the log\,\te-log\,(L/L$_{\odot}$)
diagram.  We confirm three \GD\ stars (labels a, b, and g), and 10
\DSct\ pulsators lying in the expected region of the HR-diagram.
Four of them, however, have been classified by
\citet{Uytterhoeven2011b} as hybrid pulsators although they do not
fall in the overlapping region between the \GD\ and \DSct\ stars
in our HR-diagram. One star, KIC\,04\,847\,411, was not analyzed
by \citet{Uytterhoeven2011b}. Its light curve classification
listed in Table~\ref{Table:Classification} is based on our own
analysis of the first Quarter of {\it Kepler} data.

There are ten further stars that show \DSct-like oscillations in
their light curves. Six of them have been classified by
\citet{Uytterhoeven2011b} as hybrid pulsators but do not fall in
the overlapping region in the HR-diagram. Five stars (labels i, j,
m, p, and x) are close to the hot border of the \DSct\ instability
region, two other ones (labels l and r) are distinctly hotter than
given by this border. Four stars are too cool (labels h, w) or too
evolved (labels c, f) to be hybrid, \GD\ or \DSct\ pulsators.

Four stars of our sample are reported by \citet{Uytterhoeven2011b}
to be binaries. Three of them could be identified as SB2 stars.
For the fourth one, more observations are needed to confirm its
binarity spectroscopically.

Fifteen of the analysed targets show metallicities which are lower
by more than 0.3~dex than the metallicity of the Sun. The four
stars of lowest metallicity show underabundances of about 1~dex.
Two of them, KIC\,09\,812\,351 and 12\,353\,648, have a C
abundance comparable to the solar value which might be a sign of
\LBoo\ nature. Additionally to the C abundance, this type of
variable stars is characterised by solar abundances of N, O and S.
We did not find any spectral lines of these elements in the
considered wavelength range that could be used for an abundance
determination, however. We also find that most of the analysed
stars are rather fast rotators with projected rotational
velocities above 90\,\kms.

\section{Conclusions}\label{Section:Conclusions}

We determined the fundamental parameters of 26 stars in the {\it
Kepler} satellite field of view proposed to be candidates for
\GD\,-type variables \citep{Uytterhoeven2011a}. The analysis was
done by means of the spectrum synthesis method based on the
comparison between the observed and synthetic spectra. As an
additional test of the derived \te, we computed SEDs by using
photometry from literature and determined \te\ by fitting
solar-composition \citet{Kurucz1993} model fluxes to the
photometric data.

A comparison of the results from the different methods was made.
Besides some outliers, where the reasons can be explained, the
\te\ derived from the spectrum analysis shows a good overall
agreement with the values given in the KIC. For the hottest star
of our sample, the KIC value appears to be underestimated. This
agrees with our finding in Paper\,I that the \te\ given in the KIC
are in general too low for the hotter stars because the
interstellar reddening was not properly taken into account. The
\te\ following from the SED fitting are systematically lower. This
can be explained by the interstellar reddening. Our correction for
this effect by using the equivalent widths of the interstellar
Na\,D lines to derive $E(B$$-$$V)$, improves the situation
although in some cases the resulting \te\ is found to be slightly
overestimated. The accuracy of the values for \logg\ and $[M/H]$
in the KIC is rather poor. An uncertainty of $\pm$0.5\,dex is
stated in the catalogue for both parameters, in some cases we also
find larger deviations from our analysis so that the values given
in the catalogue are not suited to check for the quality of our
findings.

The spectroscopically derived fundamental parameters allow us to
place the stars in a HR-diagram and to compare their location with
the classification made by \citet{Uytterhoeven2011b} based on the
oscillation frequencies found in the {\it Kepler} light curves. In
the result, we observed most of the stars in a relative compact
region of the diagram that reaches from the cool edge of the
\DSct\ instability strip to a region left of its hot border. For
all of the six outliers that are too cool or too evolved to fall
into the \DSct\ instability region we could find an explanation
either by binary nature or insufficient convergence of the
parameter determination.

We find three stars (labels a, b, and g) out of the 14 stars that
show oscillations in their light curves typical for \GD\ or
\GD-\DSct\ hybrid pulsators that certainly fall into the \GD\
range of the diagram. Ten further stars are found to be located in
one of the \DSct\ regions of the HR-diagram, six of them show
\DSct\ and in four \DSct\ and \GD-typical oscillations co-exist.
This shows that oscillations with periods in the \GD\ range are
much more common among the \DSct\ stars than described by the
theoretical \GD\ instability region. This finding is in agreement
with \citet{Grigancene2010} who investigated a sample of 234 {\it
Kepler} stars and found a significant number of hybrid pulsators,
whereas theory predicts the existence of hybrids in only a small
overlapping region of the instability strips.

We find seven stars close to the left or left of the blue edge of
the \DSct\ instability strip calculated for fourth radial overtone
pulsations. Only two of these stars show \DSct-like oscillations
but five of them show oscillations with periods in the \DSct\ and
in the \GD\ range. Similar to our results, \citet{Grigancene2010}
found a significant number of stars showing \DSct\ or \DSct-\GD\
oscillations which are hotter than predicted by the theoretical
blue edge of the \DSct\ instability strip calculated for fourth
radial overtone pulsations.

We found four stars with very low metallicities in the -1~dex
range. Two of them have about solar C abundance which could be a
sign of \LBoo\ nature. Both stars show \DSct-like but no
\GD-typical oscillations. Thus we did not find any hint for a
relationship between the \LBoo\ stars and \GD-type variability in
our sample.

\section*{acknowledgements} The research leading to these results
has received funding from the European Research Council under the
European Community's Seventh Framework Programme
(FP7/2007--2013)/ERC grant agreement n$^\circ$227224 (PROSPERITY).
This research has made use of the SIMBAD database, operated at
CDS, Strasbourg, France.

\label{lastpage}

\end{document}